\documentclass[conference]{IEEEtran}
\IEEEoverridecommandlockouts
\usepackage{cite}
\usepackage[numbers,sort&compress]{natbib}

\usepackage[utf8]{inputenc}
\usepackage{amsmath,amssymb}
\usepackage{graphicx}
\usepackage{bm}
\usepackage{enumerate}
\usepackage{comment}
\usepackage{xcolor}
\usepackage{url}
\usepackage{color,soul}
\usepackage{subcaption}
\usepackage{float}
\usepackage{booktabs} 

\definecolor{light-gray}{gray}{0.8}

\usepackage{amsmath,amssymb,amsfonts}
\usepackage{algorithmic}
\usepackage{graphicx}
\usepackage{textcomp}
\usepackage{xcolor}
\usepackage{subcaption}

\def\BibTeX{{\rm B\kern-.05em{\sc i\kern-.025em b}\kern-.08em
    T\kern-.1667em\lower.7ex\hbox{E}\kern-.125emX}}

\makeatletter
\newcommand{\linebreakand}{%
  \end{@IEEEauthorhalign}
  \hfill\mbox{}\par
  \mbox{}\hfill\begin{@IEEEauthorhalign}
}
\makeatother

\begin{document}

\title{Harnessing Earnings Reports for Stock Predictions: A QLoRA-Enhanced LLM Approach\\}

\author{

\small 

\begin{tabular}[t]{c@{\extracolsep{8em}}c} 

1\textsuperscript{st} Haowei Ni\textsuperscript{*} & 2\textsuperscript{nd} Shuchen Meng \\
\textit{Columbia University} & \textit{Central University of Finance and Economics} \\
New York, USA & Beijing, China \\
Corresponding author: hn2339@caa.columbia.edu & scmeng19@163.com \\

\\

3\textsuperscript{rd} Xupeng Chen & 4\textsuperscript{th} Ziqing Zhao \\
\textit{New York University} & \textit{Cornell University} \\
New York, USA & Ithaca, USA \\
xc1490@nyu.edu & zz8052@cornell.edu \\

\\

5\textsuperscript{th} Andi Chen & 6\textsuperscript{th} Panfeng Li \\
\textit{Independent Researcher} & \textit{University of Michigan} \\
Beijing, China & Ann Arbor, USA \\
real.andichen@gmail.com & pfli@umich.edu \\

\\

7\textsuperscript{th} Shiyao Zhang & 8\textsuperscript{th} Qifu Yin \\
\textit{Cornell University} & \textit{Columbia University} \\
Ithaca, USA & New York, USA \\
sz566@cornell.edu & qy2213@columbia.edu \\

\\

9\textsuperscript{th} Yuanqing Wang & 10\textsuperscript{th} Yuxi Chan \\
\textit{New York University} & \textit{Rutgers University} \\
New York, USA & New Jersey, USA \\
wangyq@wangyq.net & yoccichan.0604@rutgers.edu \\

\end{tabular}
}
\maketitle

\begin{abstract}
Accurate stock market predictions following earnings reports are crucial for investors. Traditional methods, particularly classical machine learning models, struggle with these predictions because they cannot effectively process and interpret extensive textual data contained in earnings reports and often overlook nuances that influence market movements. This paper introduces an advanced approach by employing Large Language Models (LLMs) instruction fine-tuned with a novel combination of instruction-based techniques and quantized low-rank adaptation (QLoRA) compression. Our methodology integrates `base factors', such as financial metric growth and earnings transcripts, with `external factors', including recent market indices performances and analyst grades, to create a rich, supervised dataset. This comprehensive dataset enables our models to achieve superior predictive performance in terms of accuracy, weighted F1, and Matthews correlation coefficient (MCC), especially evident in the comparison with benchmarks such as GPT-4. We specifically highlight the efficacy of the \texttt{llama-3-8b-Instruct-4bit} model, which showcases significant improvements over baseline models. The paper also discusses the potential of expanding the output capabilities to include a `Hold' option and extending the prediction horizon, aiming to accommodate various investment styles and time frames. This study not only demonstrates the power of integrating cutting-edge AI with fine-tuned financial data but also paves the way for future research in enhancing AI-driven financial analysis tools.


\end{abstract}


\section{Introduction}
In the realm of stock market prediction, relying solely on historical data to predict stock market directions has proven to be inadequate. Traditional technical indicators such as moving averages and exponential moving averages (EMAs) are frequently insufficient for accurate forecasting, especially when the market is influenced by significant events that disrupt the financial landscape. Such events include routine financial disclosures like earnings reports, which are crucial for understanding the performance of specific companies. These reports are not only pivotal but also often voluminous and time-consuming for investors to analyze comprehensively. Moreover, the plethora of metrics from balance sheets, income statements, and cash flow statements can be overwhelming and difficult to understand for those without a financial background, yet investors need this knowledge to guide their investment choices. Additionally, rare, unpredictable `black swan' events, such as the COVID-19 pandemic, can dramatically influence market behavior.

The limitation of traditional methods is their inability to account for the complex, dynamic nature of financial markets, which are often influenced by a myriad of factors beyond historical price movements. Moreover, traditional machine learning models typically struggle with predictions based on extensive, unstructured textual data like earnings reports, as they fail to capture the depth and implications of financial narratives effectively. Investors are increasingly recognizing the need for more sophisticated, reliable prediction methods that not only analyze past data but also adapt to new information and market conditions in real-time. This need becomes particularly evident during earnings seasons when companies release their quarterly financial results. These periods often witness heightened market volatility as market participants react to the new earning reports. Many investors and traders utilize options strategies to capitalize on anticipated price movements following earnings announcements. This practice, commonly known as ``betting on earnings," allows traders to potentially profit from significant price swings without directly owning the underlying stock. On the other hand, some investors use earnings reports to inform their hedging strategies, adjusting their positions to protect against potential losses or to lock in gains. The complexity of earnings reports, combined with their potential to trigger substantial market reactions, creates a double-edged sword for investors. Those who can quickly and accurately interpret these reports may gain a significant competitive edge in the market. However, the sheer volume of information and the rapidity of market reactions present formidable challenges. Many investors struggle to make timely, well-informed decisions in this fast-paced environment, potentially missing out on opportunities, making sub-optimal choices, or experiencing significant losses due to information overload or analysis paralysis.

The rapid growth and development of Artificial Intelligence have created an ideal environment for the emergence of Large Language Models (LLMs), which are transforming the field of natural language processing in many areas\cite{lai2024adaptive,ruan2024twitter,zhu2021twitter} and have sparked a great deal of interest in the diverse applications of these models. There are already established works applying advanced modeling techniques, including deep learning and LLMs, in finance. For instance, Zhang et al. \cite{zhang2023enhancing} fine-tuned an LLM using Retrieval Augmented Generation (RAG) for sentiment analysis, showcasing the use of advanced AI to enhance financial data interpretation. Similarly, Koa et al. \cite{Koa_2024} introduced the Summarize-Explain-Predict (SEP) framework, which employs self-reflective agents and Proximal Policy Optimization (PPO) to fine-tune LLMs for explainable stock predictions. Additionally, the FinBERT-LSTM model, which combines the financial-domain-focused capabilities of FinBERT with the sequential data processing strength of LSTM, demonstrates superior predictive accuracy in forecasting stock prices based on financial news sentiment analysis over traditional LSTM and DNN models \cite{gu2024predicting}. These advancements highlight the significant role of sophisticated AI in enhancing the accuracy and efficiency of financial market analysis by effectively capturing complex patterns and dependencies within the data.

Building upon these advances, our paper introduces a novel approach by fine-tuning a Large Language Model (LLM) to analyze earnings reports alongside external factors such as overall market index performance and analysts' grades. This method aims to provide investors with more reliable predictions of next-day stock performance following earnings announcements. Unlike existing models that primarily focus on historical data or limited real-time indicators, our approach leverages a comprehensive set of data inputs, enhancing the predictive accuracy and relevancy of financial forecasts. This new direction not only addresses the shortcomings of traditional forecasting methods but also capitalizes on the strengths of AI to synthesize complex datasets into actionable insights, significantly aiding investors in navigating the complexities of post-earnings stock movements.

\section{Data Collection and Preprocessing}

\subsection{Dataset Curation}

Our study focused on a comprehensive dataset encompassing 501 companies listed in the S\&P 500. The selection of 501 companies, instead of the traditional 500, accounts for occasional changes in the index due to mergers, acquisitions, or other financial adjustments that may temporarily expand the list. We utilized the API from Financial Modeling Prep to acquire this extensive financial data. The data collection process was meticulous, incorporating several key factors to ensure a robust and multifaceted analysis:

\begin{itemize}
    \item Market Performance Metrics: For each company, the performance data for the past week from major market indices (SPY, QQQ, and DOW), along with the company's own stock performance, was collected and aligned with the quarterly earnings date.
    \item Analyst Grades: A comprehensive list of stock upgrades and downgrades was compiled from various analysts for the months preceding each earnings date, providing insights into market sentiment and expert opinions.
    \item Earnings Surprises: Instances where companies either beat or missed their estimated Earnings Per Share (EPS) were identified by comparing actual EPS results with estimates.
    \item Financial Metrics Growth: This part includes a comparative analysis of financial metrics, contrasting the quarter of the earnings report with the same quarter from the previous year. This encompasses key data points typically presented in earnings reports, including:
    \begin{itemize}
        \item Income statement metrics
        \item Balance sheet figures
        \item Cash flow statement data
    \end{itemize}
    \item Earnings Transcripts: Full earnings call transcripts were collected for each company to capture qualitative data and management insights.
    \item Output Label: For the study’s outcome metric, we calculated the next day's stock performance after the earnings announcement. If the opening price is less than the closing price, we label this as `Long'; otherwise, it is labeled as `Short'. This measure serves as the dependent variable in our predictive models.
\end{itemize}

\begin{figure*}[!htbp]
    \centering
    \includegraphics[width=\textwidth]{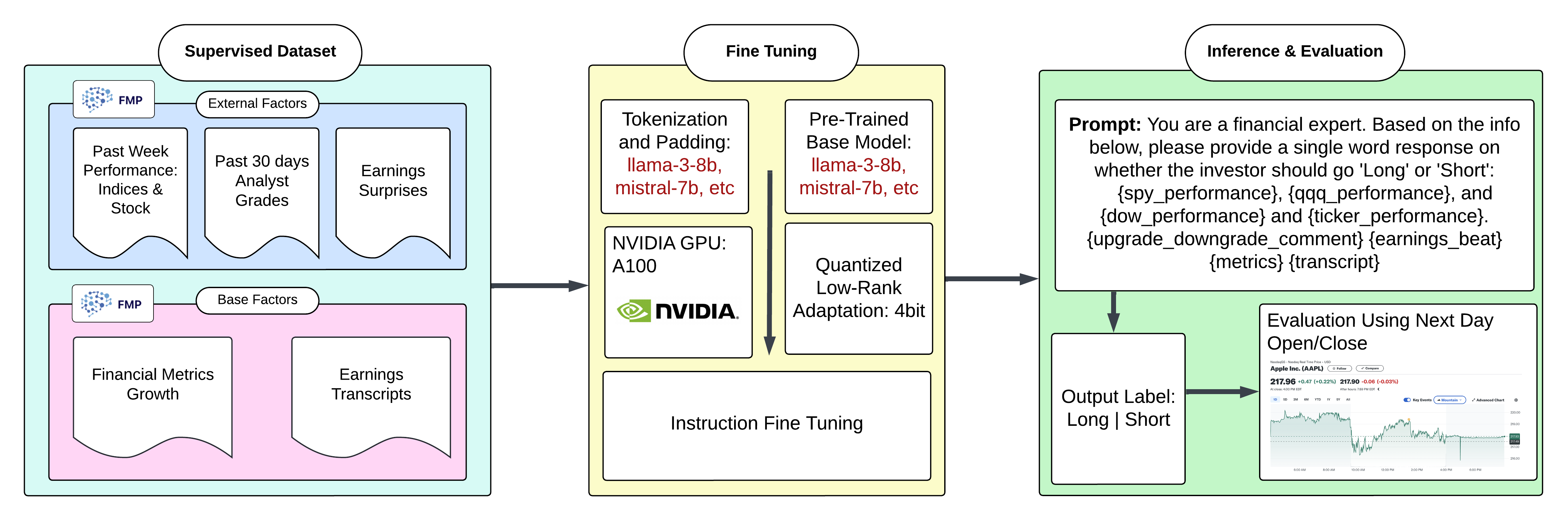}
    \caption{Overview of the framework includes dataset setup, instruction-based fine-tuning, QLoRA compression, and evaluation with designated prompts and outputs.}
    \label{fig:framework}
\end{figure*}

\subsection{Textualization and Tokenization}
In this phase, we transformed our numerical and categorical data into natural language text, enhancing the dataset's compatibility with our large language model. This process of textualization involved converting raw financial data into descriptive sentences, making the information more contextual and easier for the model to process.

\subsubsection{Market Performance Metrics}
The percentage changes of major indices and individual stocks were converted into descriptive statements.

\begin{itemize}
    \item \textbf{Original data:} ``SPY -1.5\%"
    \item \textbf{Textualized form:} ``In the past week, SPY went down by 1.5\%"
\end{itemize}

This transformation was applied consistently to SPY, QQQ, DOW, and the specific company ticker's performance data.

\subsubsection{Analyst Grades}
The analyst grades over the past 30 days were aggregated to determine the most frequent recommendation, which was then transformed into a descriptive statement:

\begin{itemize}
    \item \textbf{Original data:} Multiple analyst grades over 30 days
    \item \textbf{Aggregated data:} Most frequent grade (e.g., ``Buy")
    \item \textbf{Textualized form:} ``In the past 30 days, most grading companies suggest buying this stock"
\end{itemize}

\subsubsection{Earnings Surprises}
The earnings beat/miss data, originally consisting of actual and estimated EPS values, were transformed into comparative statements. For instance:

\begin{itemize}
    \item \textbf{Original data:} ``Actual EPS: \$2.10, Estimated EPS: \$1.95"
    \item \textbf{Textualized form:} ``The company's reported earnings per share (EPS) were 7.69\% higher than the analysts' consensus estimates"
\end{itemize}

\subsubsection{Financial Metrics Growth}
Year-over-year growth ratios for various financial metrics were converted into descriptive sentences. For example:

\begin{itemize}
    \item \textbf{Original data:} ``growthNetIncome": 0.046
    \item \textbf{Textualized form:} ``Compared to the same quarter last year, Net Income grew by 4.6\%"
\end{itemize}

Following the removal of rows with missing values, the resultant dataset comprised 8,556 rows, ensuring a comprehensive and complete dataset for the LLM training. The average token count per row is 12,375, demonstrating the substantial content size processed per instance. The minimum token count observed is 1,272, while the maximum reaches 54,394. These statistics highlight the wide range of text lengths the model must handle, which is crucial for understanding both the diversity of the dataset and the robustness required of the model. Two distinct datasets, named Base and Full, were prepared to assess the impact of various features on model performance. The Base dataset focuses exclusively on financial metrics growth and earnings transcripts, providing a baseline to evaluate internal company performance. The Full dataset includes additional text features such as overall market performance, analyst grades, and earnings estimates, allowing for an analysis of how external factors influence the model’s effectiveness. This bifurcation aims to determine the relative impact of internal versus external data on predictive accuracy.

\section{Method}
\subsection{Framework}
The architecture of our model, detailed in Figure 1, is meticulously designed to leverage both fundamental and external financial factors for making predictive analyses. Initially, a supervised dataset was constructed by integrating `base factors', which include financial metric growth and earnings transcripts, with `external factors' such as the previous week's performance of market indices, the stock itself, analyst grades, and earnings surprises. This dataset facilitated the comprehensive training of our models. Several pre-trained models were selected for further optimization to fit the specific demands of financial forecasting. The models were compressed using QLoRA, a technique designed to maintain original performance while enabling efficient deployment on resource-constrained devices, and then fine-tuned with targeted instructions to enhance their predictive capabilities.

During the inference and evaluation phase, a structured approach was adopted, using clearly defined prompts that led to binary outputs. The real-world efficacy of our models was rigorously assessed against actual market data, employing metrics such as accuracy, weighted F1 score, and the Matthews correlation coefficient (MCC). 

\begin{table*}[ht]
    \centering
    \caption{Performance comparison of different models}
    \label{tab:model_comparison}
    \small 
    \begin{tabular}{@{}lcccccc@{}}
        \toprule
        \textbf{Model} & \multicolumn{3}{c}{\textbf{Base}} & \multicolumn{3}{c}{\textbf{Full}} \\
        \cmidrule(r){2-4} \cmidrule(r){5-7}
        & \textbf{Accuracy} & \textbf{Weighted F1} & \textbf{MCC} & \textbf{Accuracy} & \textbf{Weighted F1} & \textbf{MCC} \\
        \midrule
        ChatGPT 4.0    & 0.363 & 0.482 & 0.023 & 0.494 & 0.512 & 0.031 \\
gemma-7b-4bit & 0.541 & 0.468 & 0.135 & 0.542 & 0.442 & 0.178 \\
Phi-3-medium-4k-instruct & \textbf{0.559} & 0.469 &  \textbf{0.224}  & 0.560  & 0.471 &  \textbf{0.227}  \\
Phi-3-mini-4k-instruct & 0.548 & 0.478 & 0.154 & 0.557 & 0.494 & 0.175 \\
mistral-7b-4bit & 0.556 & \textbf{0.556} & 0.112 & 0.550 & 0.497 & 0.122 \\
mistral-7b-instruct-4bit & 0.549 & 0.472 & 0.168 & 0.534 & 0.532 & 0.070 \\
llama-3-8b-4bit & 0.534 & 0.533 & 0.069 & 0.541 & 0.535 & 0.087 \\
mistral-7b-4bit & 0.542 & 0.497 & 0.114 & 0.544 & 0.536 & 0.089 \\
llama-3-8b-Instruct-4bit & 0.550 & 0.533 & 0.104 & \textbf{0.573} & \textbf{0.565} & 0.154 \\
        \bottomrule
    \end{tabular}
\end{table*}

\subsection{Instruction Fine Tuning}
Instruction fine-tuning is a technique used to enhance the performance of large language models by training them to comprehend and execute specific, structured instructions, especially useful in domains where precise and actionable outputs are needed from complex input data \cite{xiong2024large,mo2024fine}. The process involves three key steps: First, creating an instruction-following dataset consisting of paired instructions and their corresponding expected responses, which act as labels for investor actions such as `Long' or `Short'. This dataset includes relevant financial data points tailored to reflect real-world scenarios the model will encounter. Second, the model is fine-tuned using this dataset, where it learns to generate the expected responses based on the provided instructions, adapting its pre-existing capabilities to the specific task of making investment predictions \cite{chen2023hadamard,zhang2024pruning}. Finally, the generated outputs are mapped back to the predefined labels to ensure the model’s responses are accurate and applicable to practical financial decision-making settings. 

\subsection{QLoRA}
Quantized Low-Rank Adaptation (QLoRA) \cite{dettmers2023qlora} represents an innovative approach in the field of model optimization and efficiency enhancement, particularly for large pre-trained models such as transformers. This technique leverages the principles of Low-Rank Adaptation (LoRA) \cite{hu2021lora} while incorporating quantization strategies to reduce the computational demand and memory usage of the model adaptations. In our project, we implemented the 4-bit version of QLoRA to fine-tune and deploy a large language model. The primary goal of using 4-bit QLoRA was to compress the model's adaptive parameters without significantly compromising its performance, allowing the model to maintain a balance between efficiency and efficacy, making it suitable for environments with stringent resource constraints.

Quantization in QLoRA involves reducing the precision of the low-rank matrices from the standard 32-bit floating points to just 4 bits. This drastic reduction in bit depth leads to a significant decrease in the memory footprint of the model, which is especially beneficial given the lengthy nature of our prompts or tokens that require processing. The 4-bit quantization not only minimizes the storage space but also accelerates the computation, as operations with lower-bit numbers are computationally less expensive. These features make 4-bit QLoRA an essential tool for optimizing the deployment and operational efficiency of advanced AI models in resource-limited settings.

\section{Evaluation}
\subsection{Model Training}
In this study, a comprehensive fine-tuning protocol was implemented across various advanced pre-trained models to optimize their performance for specialized tasks in financial analytics. Models such as ``gemma-7b-4bit'', ``Phi-3-medium-4k-instruct'', ``Phi-3-mini-4k-instruct'', ``mistral-7b-4bit'', ``llama-3-8b-4bit'', and ``llama-3-8b-Instruct-4bit'' were fine-tuned for one epoch using a learning rate of $2e-4$. Efficiency in training was achieved by carefully configuring the batch size and gradient accumulation to manage computational resources effectively. Specifically, training parameters were adjusted to allow for significant gradient accumulation, enhancing training stability without requiring substantial memory increases. A warm-up phase consisting of five steps was integrated to gradually adjust the learning rates, which helped stabilize the pre-learned weights of the models.

Optimization utilized the adamw\_8bit'', an $8$-bit variant of the AdamW optimizer, striking a balance between performance and memory efficiency. Weight decay was set at $0.01$ to regularize the training process and prevent overfitting \cite{wu2024surveying}. The learning rate scheduler was configured to a linear'' setting to ensure a steady decrease in the learning rate as training advanced. Models were capable of processing inputs up to a maximum sequence length of $25,000$, allowing extensive financial texts to be analyzed comprehensively. This structured training approach was designed to enhance models' predictive accuracy and operational efficiency, capitalizing on their unique architectural strengths.

\subsection{Baseline Models}
\subsubsection{ChatGPT} GPT-4.0, an advanced iteration of the Generative Pre-trained Transformer series, was utilized through OpenAI's API, ensuring robust and state-of-the-art performance benchmarks. This model is renowned for its deep learning capabilities across a wide range of language-based tasks.

\subsection{Performance Analysis}
\subsubsection{Evaluation Metrics}
To assess the performance of the fine-tuned models, three key metrics were employed: accuracy, weighted F1 score, and the Matthews Correlation Coefficient (MCC). Accuracy measures the overall correctness of predictions, providing a straightforward metric of performance across all classes. The weighted F1 score, which considers both precision and recall, is particularly useful in situations where class imbalance might affect the model's evaluation. Lastly, the Matthews Correlation Coefficient (MCC) offers a comprehensive measure of the model's quality, capturing true and false positives and negatives in a single coefficient. This combination of metrics ensures a robust analysis of the models' predictive capabilities.

\begin{figure}[!htbp]
    \centering
    \includegraphics[width=0.45\textwidth]{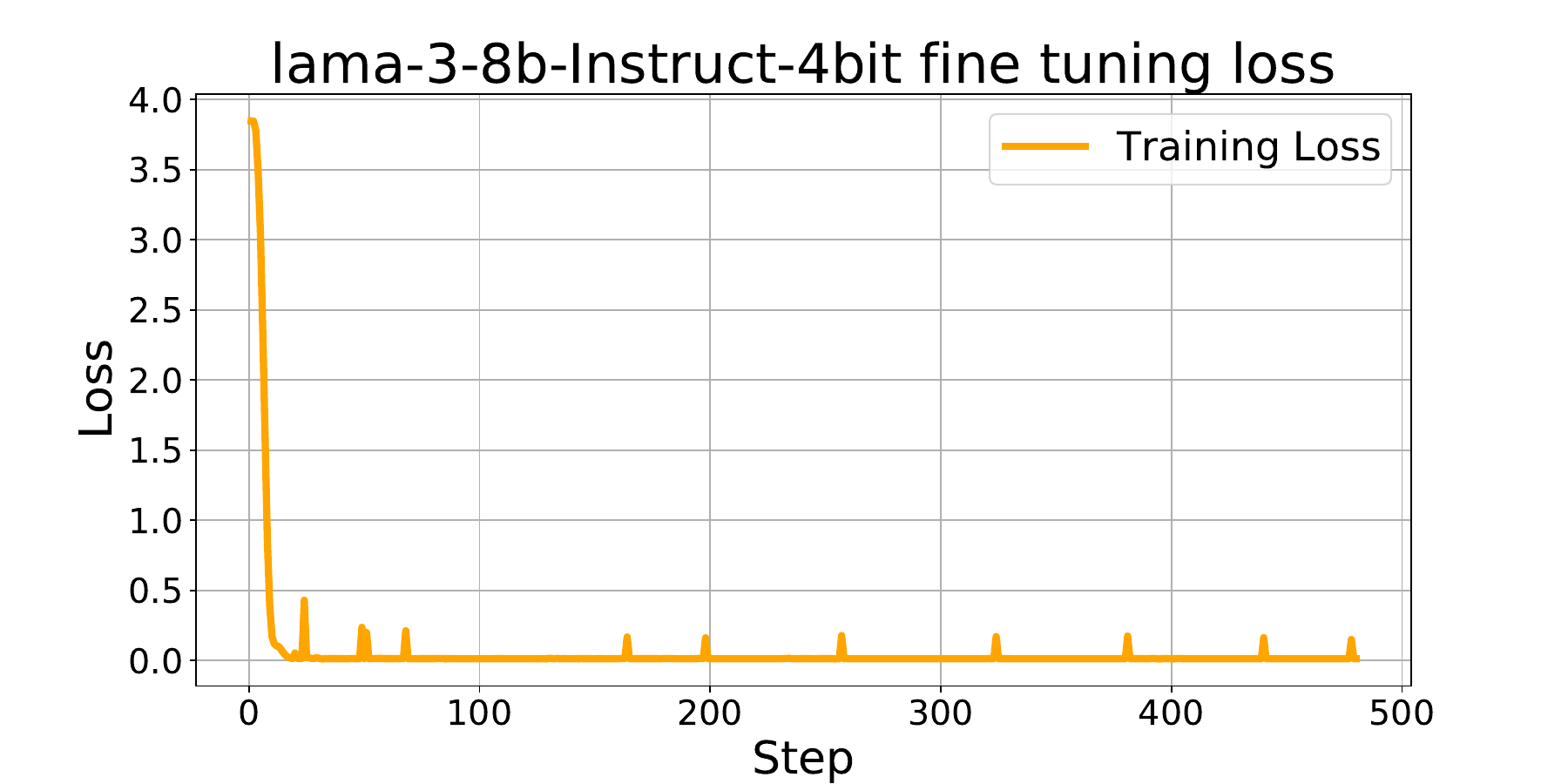}
    \caption{Training loss curve during fine-tuning of Llama 3-8b-Instruct model}
    \label{fig:framework}
\end{figure}

\subsubsection{Comparative Analysis of Model Performance}
The performance of various fine-tuned models was assessed against a benchmark established by GPT-4, utilizing both the `Base' dataset, which included only base factors, and the `Full' dataset, enriched with external factors. While the `Base' dataset provided insights into intrinsic performance, more emphasis was placed on the results from the `Full' dataset to evaluate the comprehensive capability of the models when supplemented with additional contextual information. Among the models evaluated, the llama-3-8b-Instruct-4bit, a highly optimized version of a large language model tailored explicitly for instruction-based tasks, demonstrated superior performance in the `Full' dataset. It achieved 16\% higher accuracy and 10\% better Weighted F1 compared to GPT-4, indicating its effectiveness in handling complex, multifaceted financial data. This model leverages instruction-based tuning to enhance its applicability in specific contexts, making it particularly adept at financial forecasting. This comprehensive analysis underscores the significant advancements in model accuracy and reliability when incorporating both base and external factors into financial forecasting. In addition to comparing the final performance metrics of the models, the training dynamics were also analyzed to understand the models' learning behavior throughout the fine-tuning process. A significant observation was the decrease in fine-tuning loss, as illustrated in Figure 2. This graph demonstrates a consistent reduction in loss across training steps, which indicates effective learning and adaptation of the models to the task. The smooth descent in loss underscores the efficiency of the training regimen and parameter settings chosen for the fine-tuning phase. This steady improvement in loss metrics not only validates the training strategy but also highlights the stability of the models during training, reflecting their robustness in capturing the complexities of financial data.

\section{Conclusion and future work}

This research demonstrated the efficacy of employing advanced machine learning techniques, specifically instruction-based fine-tuning coupled with QLoRA compression, to enhance the predictive accuracy of stock market forecasts post-earnings announcements. By systematically integrating both `base factors', such as financial metric growth and earnings transcripts, with `external factors', including past week market indices and stock performance alongside analyst grades, this study crafted a comprehensive dataset that enabled the fine-tuned models to capture the multifaceted nature of financial data effectively. The instruction-based fine-tuning process, tailored to the nuanced demands of financial forecasting, significantly augmented the capabilities of our models. This was evidenced by the substantial improvements in performance metrics, particularly when evaluated on the `Full' dataset, which included a broader range of contextual information compared to the `Base' dataset. The LLM, particularly the llama-3-8b-Instruct-4bit model, displayed remarkable performance enhancements, outstripping traditional approaches such as GPT-4 in terms of accuracy, weighted F1, and MCC. Additionally, the effectiveness of fine-tuning transformer-based large language models suggests a promising foundation for future research in various domains. Furthermore, the application of QLoRA provided an efficient means of deploying these advanced models on resource-constrained devices, maintaining high levels of predictive accuracy while minimizing operational demands. This aspect of model compression is crucial for real-time financial applications, where timely and effective decision-making is essential.

Looking ahead, this study acknowledges the limitation of using a binary `Long' and `Short' output in its predictive model, which may not fully capture the nuances of investor behavior in real-world scenarios. Recognizing that investors often adopt a `Hold' stance as a cautious or neutral position, future work will aim to include this option in the model's output capabilities. Additionally, considering the diverse investment strategies that focus on longer-term trends, expanding the prediction horizon beyond the next day to encompass weekly movements could align better with the needs of various investor types. These enhancements will strive to make the model a more comprehensive tool for investment decision-making, accommodating a broader range of investment styles.

Overall, the integration of sophisticated AI techniques with a well-curated dataset has proven to be a powerful approach in financial analytics, offering substantial improvements over traditional models. This approach not only addresses the limitations of existing forecasting methods but also sets a new benchmark for the application of AI in financial decision-making. The success of this study paves the way for further explorations into the integration of more diverse data types and advanced modeling techniques in the financial domain.

\renewcommand{\bibfont}{\footnotesize}

\footnotesize{
\bibliographystyle{IEEEtran}
\bibliography{main}
}

\end{document}